# Stacked polarimeters with twisted black phosphorus

**Authors**
Yifeng Xiong[1,9], Yushu Wang[1,9], Runze Zhu[1,9], Haotian Xu[1], Chenhui Wu[1], Jin-hui Chen[2], Yang Ma[1], Yuan Liu[1], Ye Chen[1], K.Watanabe[3], T. Taniguchi[4], Mengzhu Shi[5,6], Xianhui Chen[5,6], Yanqing Lu[1], Peng Zhan[7], Yufeng Hao[1,8*], Fei Xu[1*]

**Affiliations**
[1]National Laboratory of Solid State Microstructures, College of Engineering and Applied Sciences, Jiangsu Key Laboratory of Artificial Functional Materials, and Collaborative Innovation Center of Advanced Microstructures, Nanjing University, Nanjing 210023, China
[2]Institute of Electromagnetics and Acoustics, Xiamen University, Xiamen 361005, China
[3]Research Center for Functional Materials, National Institute for Materials Science, 1-1 Namiki, Tsukuba 305-0044, Japan
[4]International Center for Materials Nanoarchitectonics, National Institute for Materials Science, 1-1 Namiki, Tsukuba 305-0044, Japan
[5]Hefei National Laboratory for Physical Sciences at Microscale and Department of Physics, and CAS Key Laboratory of Strongly-coupled Quantum Matter Physics, University of Science and Technology of China, Hefei, Anhui 230026, China
[6]CAS Center for Excellence in Quantum Information and Quantum Physics, Hefei, Anhui 230026, China
[7]School of Physics and National Laboratory of Solid State Microstructures, Nanjing University, Nanjing 210093, China
[8]Haian Institute of New Technology, Nanjing University, Haian 226600, China
[9]These authors contributed equally: Yifeng Xiong, Yushu Wang and Runze Zhu
Email: haoyufeng@nju.edu.cn; feixu@nju.edu.cn

**Abstract**
   The real-time, in-line analysis of light polarization is critical in optical communication networks, which suffers from the complex systems with numerous bulky opto-electro-mechanical elements tandemly arranged along optical path. Here, we propose a fiber-integrated polarimeter with nano-thickness by vertically stacking three two-dimensional (2D) materials based photodetection units. We demonstrate a self-power-calibrated, ultrafast, unambiguous detection of linear (LP) and circular polarized (CP) light according to the symmetry broken induced linear photogalvanic effects (LPGE) and circular photogalvanic effects (CPGE) in black phosphorous (BP) units, which are twistedly stacked to substitute traditional mechanical rotation of polarizers. As a demonstration, we achieve Hadamard single-pixel polarimetric imaging by the polarimeter to recognize the polarization distributions, showing potential in high-speed polarization-division-multiplexed imaging and real-time polarized endoscopy. This work provides a new strategy for next-generation ultracompact optical and optoelectronic systems, and guides a way for developing high-resolution arrayed devices with multifunctional pixels.

## Introduction

Polarization, as one of the fundamental characteristics in optical communication, imaging, navigation, sensing, and almost all optics-related fields[1-3], is still challenging to real-time monitor in optical communication networks. The downsizing of the polarization analysis system is in accordance with the trend of optical and optoelectronic integration, aiming to achieve a small system volume, avoid unnecessary energy loss, and reduce production costs. Conventional polarimeters used in optical network require a series of opto-electro-mechanical elements (including lenses, prisms, polarizers, waveplates, filters, photodetectors, mechanical parts and so on) tandemly arranged along the optical path, which is difficult to minimize and integrate[4,5]. Therefore, the realization of ultracompact polarimeters requires breakthroughs in the following aspects: i) The downsizing of bulky optical devices to micrometer-scale and the compression of numerous arranged optical components to nano- or sub-micrometer thickness. ii) Effectively separation of the impact of light power and polarization states on photodetector signals. iii) Excellent polarized photodetection in communication wavelength, such as high photodetectivity, high polarization ratio, fast temporal response, and linear/circular polarization distinguishability.

To solve the problem, several polarization-sensitive photodetectors have been investigated using natural anisotropic two-dimensional (2D) materials with low crystal symmetry, including black phosphorus (BP), $ReX_2$ (X = S, Se), and MX (M = Sn, Ge; X = S, Se)[6-15]. Such filter-less functional photodetectors have nanoscale device volumes and are promising for building arrayed devices with ultrahigh pixel densities. However, these photodetectors need external powermeters to calibrate the light power to ensure that polarization is the only variable to photocurrent. Meanwhile, the polarized photocurrent overlaps within a period of polarization state change, hence the actual polarization state of light is still challenging to precisely detect using a single device. Alternatively, artificial plasmonic hybrid structures have also been widely investigated to build polarization-sensitive photodetectors[16-20]. Recently, metasurface-mediated graphene photodetectors were reported with calibration-free linear polarization distinguishability[17,18]. These artificial plasmonic structures usually require an array of nanoantenna units as additional filtering layers to accumulate the photocurrent, thus resulting in an increase in the device size and will generate unnecessary energy loss. Moreover, it is still challenging to simply distinguish the photocurrent generated by LP or CP light.

In this work, we demonstrate a nano-thick polarimeter by vertically stacking three photodetection units based on atom-thin 2D materials onto an optical fiber endface. We choose an isotropic $Bi_2Se_3$ as the calibrate unit to monitor the illumination power, which is polarization insensitive to vertical incident light and has narrow bandgap of 0.3 eV[21]. Two anisotropic BP units are twistedly stacked above for precise polarized photosensing, which have layer-dependent direct bandgap (changes from 0.3 eV for bulk crystal to 2.0 eV for monolayer), high mobility and anisotropic optoelectronic properties[11,22-24]. The asymmetrically designed electrodes in BP units break the system symmetry and lead to a LPGE and CPGE, showing a high-contrast and ultrafast self-driven photoresponse to LP and CP light. With the power calibration unit, the different polarized photocurrents generated from the twisted BP units lead to an unambiguously detection of polarized light, without any external optical components or mechanical parts. Furthermore, we demonstrate Hadamard single-pixel polarimetric imaging by the optical-fiber integrated polarimeter to recognize the spatial polarization distributions, with prospects in applications of polarization-division-multiplexed imaging and real-time polarized endoscopy. With the basis of our scheme, it is possible to build ultrahigh-density arrayed devices without sacrificing the lateral single-pixel size by stacking ultrathin multifunctional units along the optical path. It is beneficial for building next-generation ultracompact optical systems to realize diversified and complex optical and optoelectronic applications.

## Results

**Design concept.** The designed fiber-integrated polarimeter is schematically shown in **Fig. 1a**, in which three functional units are sequentially assembled onto a flat optical fiber endface. Each unit consists of a pair of asymmetrically designed electrodes for generating a self-driven effect, a layer of light-harvesting materials for functional photodetection and a layer of hexagonal boron nitride (hBN) nanoflakes for insulation and encapsulation. Since the value of polarized photocurrent of an anisotropic material is influenced by both illumination power and polarization state, we use a polarization insensitive isotropic $Bi_2Se_3$ layer at the bottom (calibration unit) to calibrate incident light power and two twistedly stacked anisotropic BP layers for precise polarized photosensing, substituting traditional mechanical rotation of polarizers. The intersection angle between the three electrodes is designed to be 60° (**Fig. 1b** and **c**) and the armchair crystal orientation of the BP is along with the corresponding electrode, presenting the high mobility directions[25], which can be validated by the Raman spectroscopy results in **Fig. 1g** and Supplementary Fig. S2. A transmission electron micrograph (TEM) of a completed stacked device is shown in **Fig. 1d**. The overall thickness of the van der Waals heterostructure is ~ 170 nm, consisting of a layer of ~ 60 nm thick $Bi_2Se_3$, two layers of ~ 40 nm thick BP and three layers of ~ 10 nm thick hBN. The high-resolution TEM images of the BP/hBN and $Bi_2Se_3$/hBN interfaces are shown in **Fig. 1(e-f)**, respectively. We can find that $Bi_2Se_3$ has a layered structure comprising quintuple layers with a thickness of 9.6 Å. The layered structure of BP and hBN can be seen clearly with interlayer distances of 5.4 and 3.3 Å, respectively. Moreover, a thin amorphous layer of surface oxidation film ($PO_x$) can also be observed at the BP/hBN interfaces, due to the atmospheric exposure during fabrication process. With this structure design, the polarization of incident light can be unambiguously detected by the different value of polarized photocurrents generated by the twisted BP units with power calibration. Moreover, we support the self-driving mode of the device through asymmetrical electrode design, which will not only reduce energy consumption, but also improve the overall performance of the device by the suppression of dark current.

**Photosensing performance of the device.** The excellent photosensing ability of a material is critical for high-performance polarized photodetection. The narrow bandgap of $Bi_2Se_3$ and BP leads to strong photoresponses to infrared illumination. At first, we investigate the photodetection performance of the isotropic $Bi_2Se_3$ unit in Supplementary Fig. S7. The relationship between illumination power $P$ and polarization insensitive photocurrent ($I_p = I_{light} - I_{dark}$) of Bi$_2$Se$_3$ can be well fitted as $I_p = mP^n$ (m and n are fitted constants), benefiting for power calibration. Then we focus on the polarized photosensing performance on BP units, as displayed in **Fig. 2**. The initial polarization state (0°) of the incident light is fixed corresponding to the armchair direction of the BP, which shows the maximum polarized photocurrent[9]. An obvious photocurrent is observed at zero bias in the inset of **Fig. 2a**, revealing that the device can work under bias-free conditions. Photoresponsivity ($R$) and photodetectivity ($D^*$) are key parameters to quantify the performance of the BP photodetector. Considering the huge difference in the dark current of the device with and without bias, it is more convincing to use $D^*$ to evaluate the photosensing performance. The $R$ and $D^*$ curves in **Fig. 2b** show near-linear curves at log-log coordinates under bias voltages, indicating the dominance of photogating effect caused by defects. While under zero bias, the photoresponsivity hardly decreases under low illumination power due to that the defects in the BP crystal are not fulfilled by photoinduced carriers. Here, under a light power of 50 nW and bias of 0.5 V, the device shows a $R$ of ~8.52 A W$^{-1}$, an external quantum efficiency ($EQE$) of 682.29%, and a $D^*$ of 2.38×10$^9$ Jones. Regarding the bias-free condition, the device shows a $R$ of ~7.09 mA W$^{-1}$, a $EQE$ of 0.57%, $D^*$ of 2.91×10$^8$ Jones. Although the device exhibits much lower $R$ or $EQE$ at zero bias than under bias, the $D^*$ value is relatively approximative due to the strong suppression of the dark current.

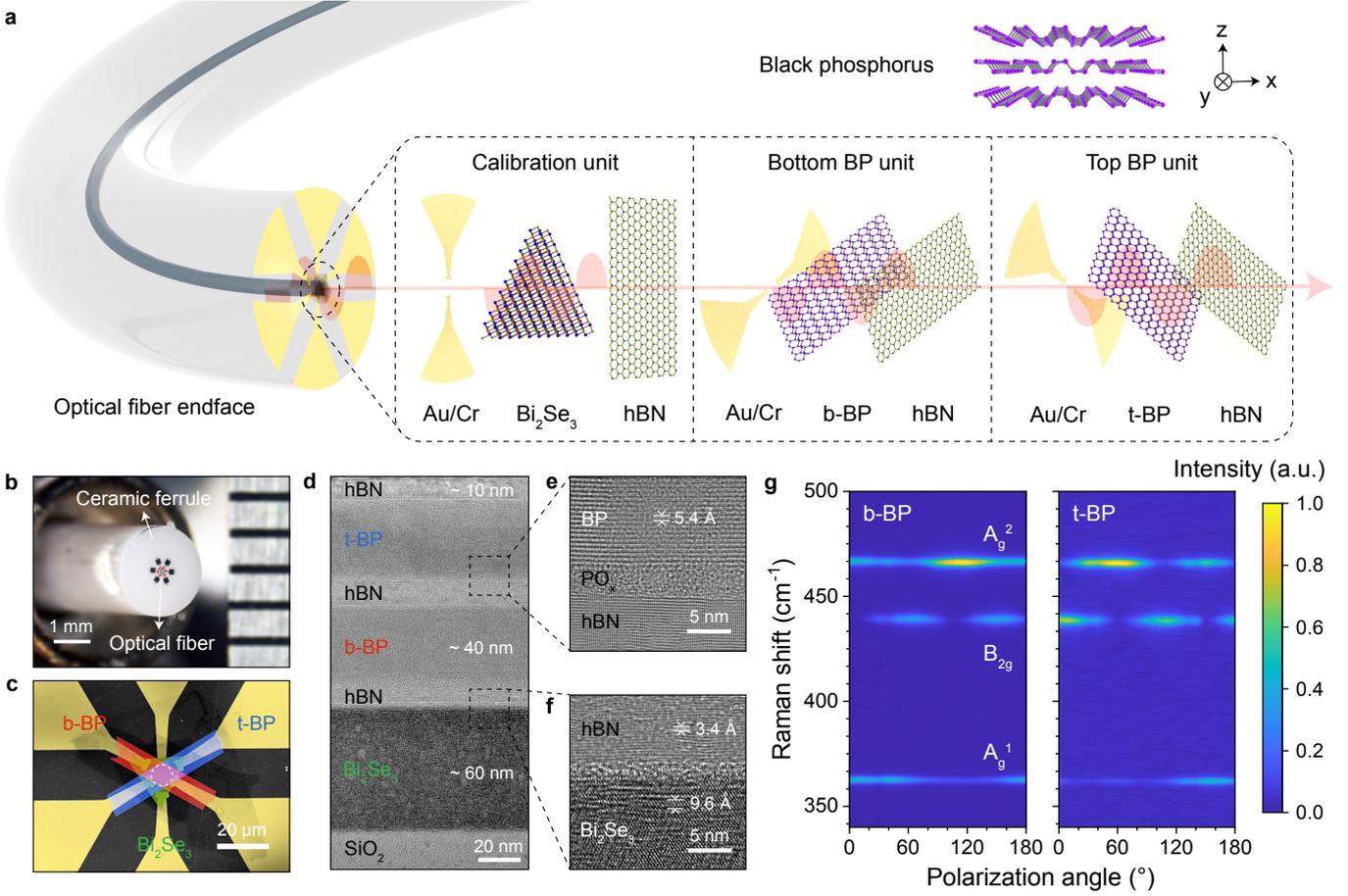

**Figure 1 Design concept of the self-driven fiber-integrated polarimeter. a**, Schematic illustration of the designed fiber-integrated polarimeter. Three functional units are stacked subsequently onto an optical fiber endface. The top-right corner is the puckered honeycomb lattice of BP, where x and y present its armchair and zigzag crystal orientations, respectively, and z presents the stack direction of its structure. **b**, Camera image of the as-fabricated device. The optical fiber is fixed in a ceramic ferrule with an outer diameter of 2.5 mm. **c**, Pseudo-color scanning electron microscope image of the stacked structures on the optical fiber endface. The white dash circle represents the fiber core area. **d**, Cross-sectional TEM image of the 2D materials van der Waals stacks. **e** and **f**, High-resolution cross-sectional TEM image showing the BP/hBN and Bi$_2$Se$_3$/hBN interfaces, respectively. **g**, Polarized Raman intensity mapping of two twisted BP nanoflakes as a function of Raman shift and polarization angle.

In addition, response time ($\tau$) and polarization ratio (PR, defined as the ratio of maximum and minimum polarization-dependent photoresponse[18]) are also crucial figures of merit to evaluate the performance of a polarization photodetector. **Figure 2c** shows the photoswitching performance of the BP unit under different bias voltages. We can find that though a larger photocurrent is obtained with an external bias voltage, the response time is much slower due to the large drift current (**Fig. 2d**). Moreover, the PR of the device under zero bias is larger than that under bias (**Fig. 2e**), benefiting for high contrast polarization distinction. **Figure 2f** shows the measured frequency response of ~1 MHz for the self-driven polarimeter. The rise and fall $\tau$ of the BP unit, measured under a 100 kHz modulation of light, are 800 ns and 850 ns, respectively. The response time of the device is the fastest among the reported BP-related photodetectors up-to-now[9,26-28], which may stem from the efficient electron-hole separation near the BP-Au interface and the fast carrier recombination within the BP crystal away from the electrode. In a brief summary, the self-driven mode of the device has greater potential in practical polarization applications, which shows a large $D^*$, a fast τ, a large PR, as well as a high on/off ratio (up to $10^3$).

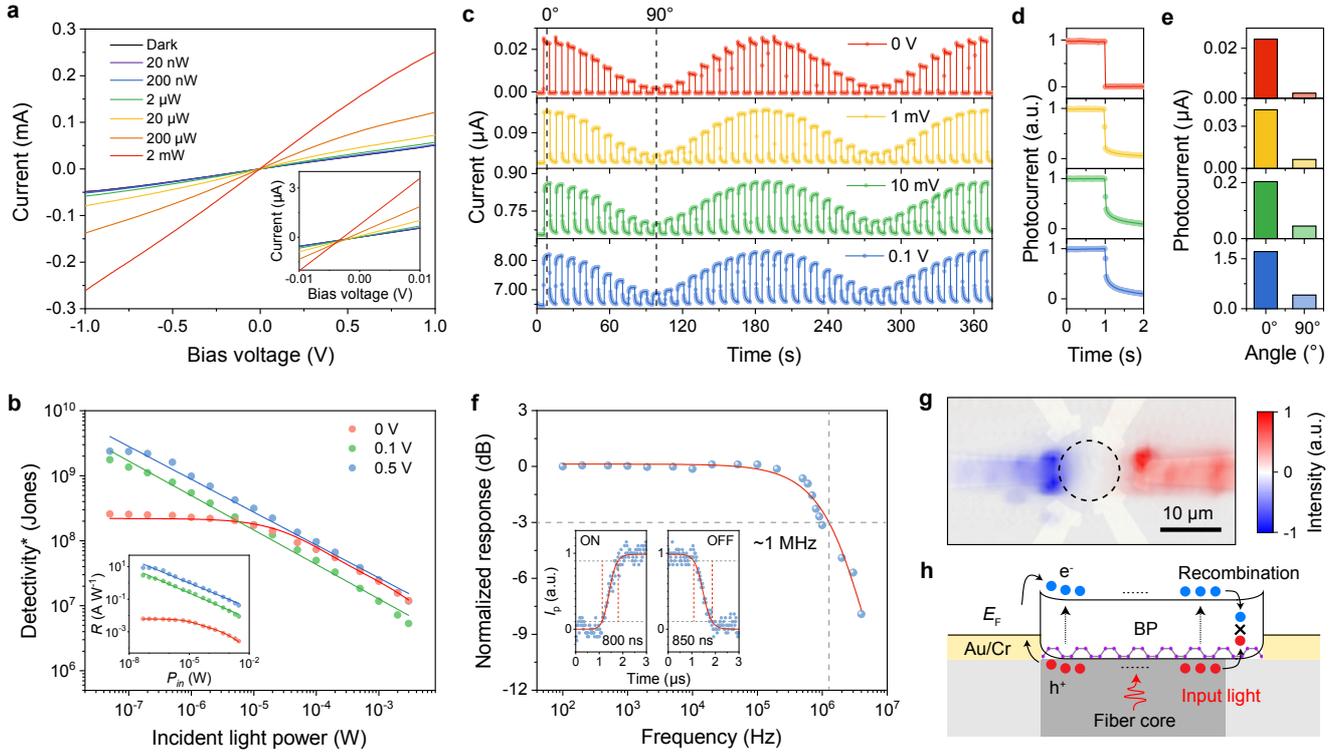

**Figure 2 Photosensing performance of the BP unit (λ = 1550 nm). a**, *I-V* curves as a function of illumination power. Inset: The enlarged view of the curves at approximately zero bias. **b**, Photodetectivity versus illuminating power under different bias voltages. The lines are fitting curves of measured data. The inset shows the corresponding photoresponsivity relating to illuminating power under different bias voltages. **c**, Bias voltage-dependent photoswitching performance under alternating dark and light illumination. The polarization angle of the incident light rotates by 10° after each light on-off cycle is completed. **d**, Half cycle of photocurrent dynamics (from **c**) under different bias voltages. **e**, Photocurrent generation versus bias voltage, which are varied for input light polarization along the x and y crystal orientations of BP. **f**, Frequency response of the device under zero bias, showing a -3 dB frequency of ~1 MHz. Inset: One cycle of photocurrent dynamics under zero bias. **g**, Photocurrent mapping of BP under zero bias. The black dash circle represents the fiber core area. **h**, Schematic of band diagram and photoexcited carriers transport under light illumination. The self-driven photosensing is based on the designed asymmetric electrodes.

To further illustrate the self-driven effect, **Fig. 2g** shows that the asymmetrical design of electrodes against the fiber core can generate spontaneous photocurrent, which can be explained by the energy diagrams in **Fig. 2h**. At the BP-Au interface, a Schottky junction is formed and leads to the formation of a built-in electric field with a 160 meV potential difference measured by the kelvin probe force microscopy (Supplementary Fig. S3). Photoinduced electron-hole pairs generated at the illumination area will diffuse to the region with a low carrier concentration. The recombination of carriers during the diffusion process results in the asymmetry of the number of carriers injected into the two electrodes, thus bringing a net photocurrent.

**Self-driven polarization analysis of the polarimeter.** As discussed above, with the help of $Bi_2Se_3$ power calibration unit, the polarization of light can be detected by measuring the photocurrent generated by BP unit. However, due to the overlap of the polarized photocurrent value within a period of polarization state change, a single BP unit cannot fully distinguish all polarization states. Therefore, we first demonstrate the LP light discernment of the polarimeter with two twisted BP units, as shown in **Fig. 3a**. The measured results can be well fitted with $I_b = a_1+b_1\cos(2\theta)$ and $I_t = a_2+b_2\cos(2\theta-2\pi/3)$, where $a_1$, $b_1$ and $a_2$, $b_2$ are related to the light power and can be calculated with the assistant of $Bi_2Se_3$ calibration unit. As we expected, the polarized photocurrent curve of t-BP and b-BP shows a 60° offset, which matches well with the polarized Raman result in **Fig. 1d**, indicating that the twist crystal orientations of b-BP and t-BP is the reason for the difference in the polarized photocurrents. Notably, the slight change of the polarization state of the light transmitted from b-BP (due to the anisotropic absorption of BP) shows no significant impact on the polarization detection of t-BP.

In addition to the LP light detection, we also demonstrated the CP photoresponses of the polarimeter. Due to the broken of inversion symmetry of a material, the illumination of circularly polarized light onto the material results in a helicity-dependent photocurrent, which is called CPGE.[29-36] In our polarimeter, the presence of anisotropic electric fields near the metal-BP contacts (origins from the Schottky barriers) will reduce the symmetry of the system and result in a net circularly polarized photocurrent[37-39]. The helicity of light was continuously modulated by a quarter-wave plate (QWP), and the initial state of the QWP (0°) was set when the fast axis of the QWP coincided with the polarization direction of incident LP light. In a 180° QWP modulation period, the light goes through linear (0°)–left circular (45°)–linear (90°)–right circular (135°)–linear (180°) polarization states, and the photocurrent is found to change periodically with the change of light helicity. The generated photocurrent in **Fig. 3b** can be quantitatively expressed as

$$J = J_C \sin(2\varphi) + J_L \cos(4\varphi + \varphi_0) + J_0,$$

where $\varphi$ is the QWP angle, $J_C$ is the amplitude of the helicity-dependent CPGE current, $J_L$ is the amplitude of the linear polarization-sensitive LPGE current with a phase shift $\varphi_0$, and $J_0$ is the polarization-independent background current[31,32]. It is obvious that CPGE currents are generated in both BP units, whose direction can be reversed by changing the light polarization from left circular polarization (LCP) to right circular polarization (RCP).

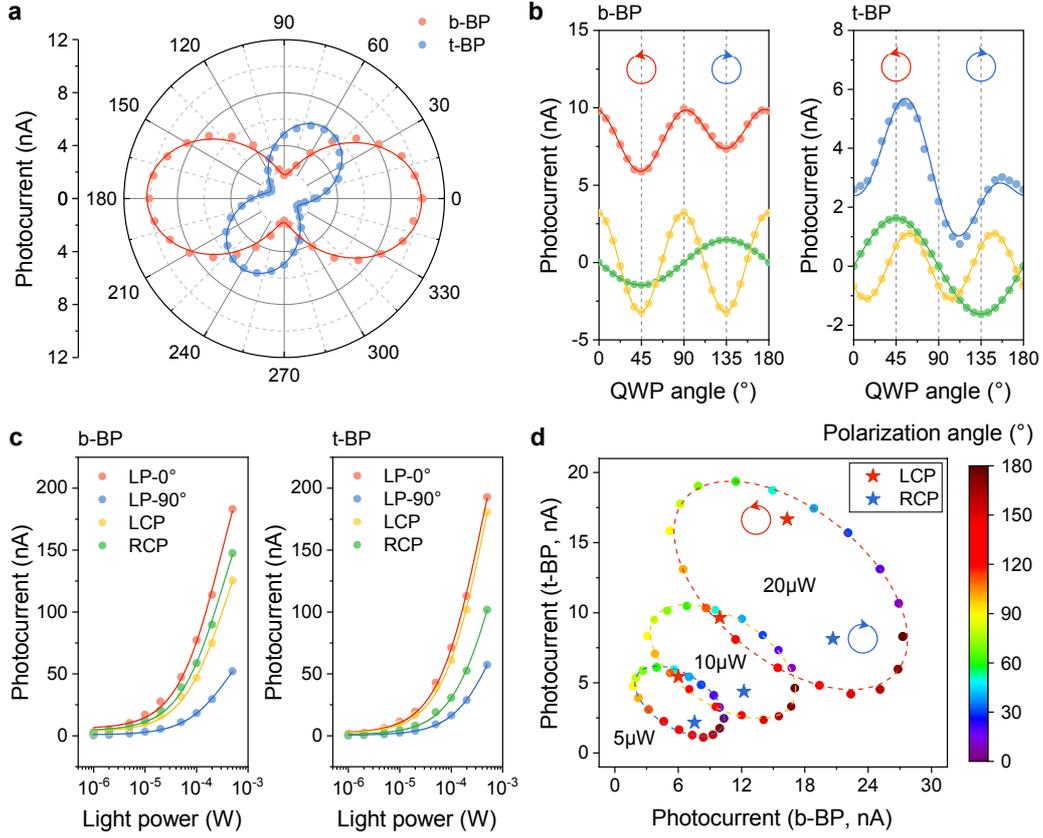

**Figure 3 Self-driven polarization analysis of the device. a**, Polar plots of the polarized photocurrent generated in twisted BP units as a function of the linear polarization angle. The circles are experimental data, and the curves are fitted results. **b**, One period of photocurrent of the twisted BP units as a function of quarter wave plate (QWP) angle (red and blue curves). The extracted CPGE (green curves) and LPGE (yellow curves) are current components of the generated circularly polarized photocurrent. The red and blue circles are experimental data and the curves are fitted results. **c**, Photocurrents of the twisted BP units as functions of light power with different polarization state of incident light. **d**, Two-dimensional plot of photocurrent generated in twisted BP layers under different polarization states. The results of linearly polarized light analysis fit well with the dark dashed ellipse.

**Figure 3c** shows the relationships between light power and generated photocurrent in BP units under different polarization states. By power calibration, we can obtain the maximum and minimum value of photocurrents in BP units during a period change of LP light states, thus establish sine equations. Therefore, when the incident light is LP, the photocurrent data from b-BP ($I_b$) and t-BP ($I_t$) forms a 2D parameter space that corresponds well to an elliptic equation:

$$(a\,I_b + b)^2 + (c\,I_b + d\,I_t + e)^2 = 1,$$

where a-e are constants related to the sine equations of b-BP and t-BP (see supplementary information for details). **Figure 3d** shows the 2D plot of the measured photocurrent in the twisted BP units with the dots colored based on the polarization state of light, which fulfill the calculated elliptic equations. As for the incident light is CP, the photocurrents generated by LCP or RCP states can be obtained by **Fig. 3c** with power calibration, and the photocurrent data is shown as the intersection point of two straight lines in the $I_b$-$I_t$ parameter space. Therefore, the polarization state of an unknown LP and CP light can be unambiguously detected using the polarimeter.

**Single-pixel infrared polarimetric imaging.** As discussed above, the proposed device has a strong polarization distinguishing ability with ~1 MHz detection bandwidth, which is promising for real-time polarimetric infrared imaging. Single-pixel imaging (SPI) is a computational imaging method that needs only a single-pixel detector instead of a pixelated array detector, showing the advantages of fast speed, high signal-to-noise ratio, and low cost[40,41]. In this context, based on the waveguide characteristics of optical fiber, our fiber-integrated polarimeter can be applied to achieve polarimetric imaging under different polarized illuminations by SPI technique.

Here, we used Hadamard single-pixel imaging (HSI) technique, which has good noise robustness and imaging quality [41]. As shown in **Fig. 4a,** a polarized 1550 nm laser beam was modulated by a digital micromirror device (DMD) to generate a series of Hadamard encoded patterns. The modulated light illuminated the target object, and the light signals were detected by the device and converted to self-driven photocurrent. One

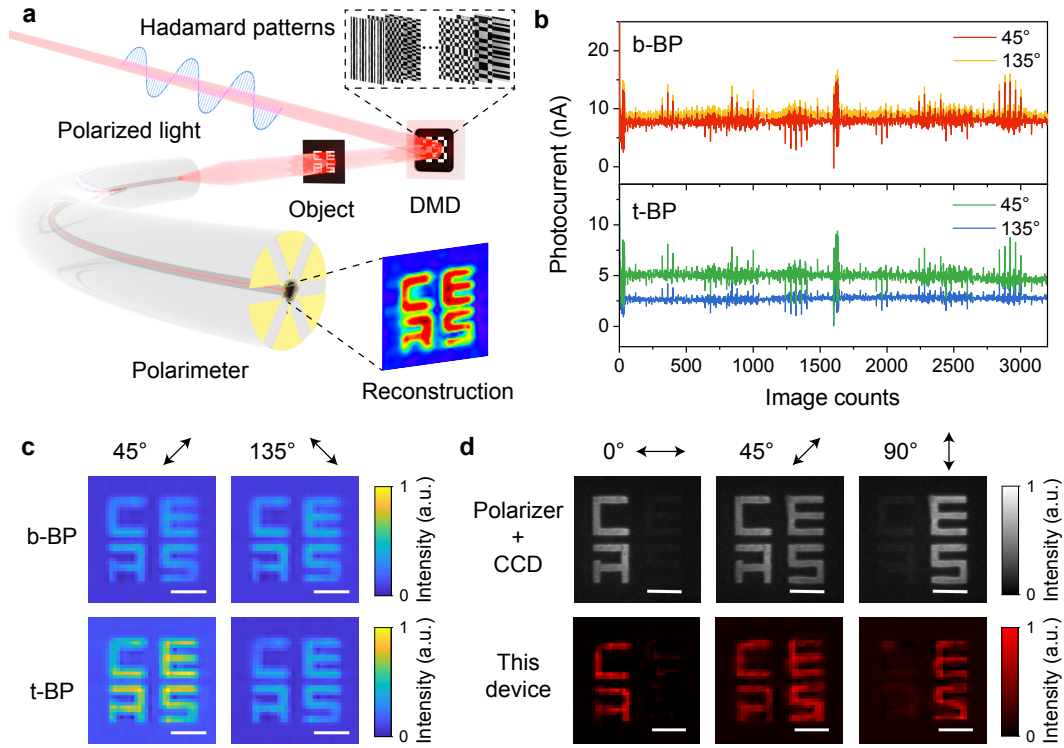

**Figure 4 Single-pixel polarimetric imaging by the device. a**, Schematic view of the experimental setup. **b**, Self-driven photocurrents generated in both the bottom and top BP layers under 45° and 135° linear polarized illuminations. **c**, Reconstructed polarimetric images of b-BP (top row of figures) and t-BP (bottom row of figures) under 45° and 135° linear polarized illuminations. The scale bars are 1 mm. **d**, Comparison of the images recorded by the CCD with a polarizer and the images reconstructed by the device under different linear polarized illuminations. The scale bars are 1 mm.

Hadamard pattern corresponds to a specific photocurrent value. The self-driven photocurrents generated in both the bottom and top BP layers under 45° and 135° linear polarized illuminations are shown in **Fig. 4b**, and the initial polarization direction (0°) is fixed along the armchair of the bottom BP. The object images can be computationally reconstructed by differential HSI technique (more details in Supplementary), and the reconstructed images under 45° and 135° linear polarized illuminations are displayed in **Fig. 4c**. The image intensity was normalized according to the photocurrent value to better show the responses of the BP detector to polarized light. By the stacking structure of BP, we can not only achieve object imaging, but also effectively distinguish the polarization of light. For example, the image intensities of the 45° and 135° polarizations are almost the same for the b-BP unit, which can be distinguished by the image obtained from the t-BP unit with strong intensity differences.

**Fig. 4d** demonstrates the capability of the device for recognizing non-uniform polarization distribution of light field, which has important applications in the field of medical treatment, astronomy, industrial inspection, etc[12,42]. The objects "CA" and "ES" were covered by two oriented tape films with a 45° difference (Supplementary Fig. S13), which can act similarly to a half-wave plate and change the polarization state of the transmitted light[43]. Different linear polarized light illuminated the object and the images reconstructed by the device (bottom row of **Fig. 4d**) are consistent with the images recorded by an infrared CCD with a polarizer (top row of **Fig. 4d**).

Optical fibers are widely used in endoscopic imaging because of their small size and flexibility. Our fiber-integrated polarimeter shows the ability to directly extract polarization information at a ~1 MHz rate, free from extra polarization modulation and analysis devices, showing the potential to realize miniaturized, low-cost, and real-time polarimetric endoscopic imaging through raster scanning and single-pixel imaging. In addition, this device configuration also shows great potential for fabricating dense detector arrays to realize pixelated polarimetric imaging. The traditional division-of-focal-plane polarimeter structure requires at least four pixels in a group to acquire four polarization-dependent intensity components (0°, 45°, 90° and 135°) of the incident light, while for our device, a single-pixel can achieve the same results. Therefore, the proposed device shows potential for high-resolution and high-speed polarimetric imaging.

## Conclusion

Integration in the direction of optical path is promising for developing ultracompact optical systems with small device volume, opening up the opportunities to simultaneously analyze or modulate multiple parameters of light. Here, by van der Waals stacking ultrathin and transparent functional units on fiber endfaces, we demonstrate a fiber-integrated polarimeter for the applications of polarization analysis and polarimetric imaging. In the stacked structure, polarization insensitive isotropic $Bi_2Se_3$ unit is used to calibrate illumination power and two twisted anisotropic BP units are used for polarized photodetection. The asymmetrically designed electrodes break the symmetry of the system and lead to LPGE and CPGE of BP. Under power calibration by $Bi_2Se_3$ unit, both the LP and CP states of incident light can be fast and unambiguously detected, according to the different polarized photoresponses generated from the twisted BP photodetectors. Furthermore, we also demonstrate single-pixel infrared polarimetric imaging by the polarimeter, showing potential for future real-time endoscopic analysis and high-speed polarimetric imaging. Notably, by stacking functional units along the optical path, a single-pixel can be functionalized while maintaining the lateral size, which is especially critical for acquiring high-resolution images while obtaining more information of light. Meanwhile, we propose a typical method to design and fabricate complex structures on optical fiber endfaces, bringing possibilities to assemble diversified integrated functional units, such as power meters[44], wavelength meters[45], and modulators[46], onto the fiber endface to realize highly integrated multifunctional fiber-integrated systems.

## Materials and methods

**Materials preparation.** We used a mechanical exfoliation method to prepare BP and hBN nanoflakes, as shown in Supplementary Fig. S1. Generally, it is difficult to directly locate the x- and y-axis directions of mechanically exfoliated BP nanoflakes under a microscope. Thus, we used an oriented mechanical exfoliation method where a BP bulk crystal with a clear crystal orientation underwent repeated mechanical exfoliation while maintaining the same crystal orientation. The oriented BP flakes were finally transferred to polydimethylsiloxane (PDMS) for the next transfer step on the optical fiber endface.

The thickness of BP nanoflakes was measured by atomic force microscopy (AFM, Oxford, Cypher S) and TEM (FEI, Talos F200X) for ~40 nm. The thicker BP nanoflakes possessed a narrower bandgap, which was beneficial for infrared light harvesting and demonstrated stronger light absorption. However, the nanoflakes cannot be too thick, which would lead to a large dark current, poor performance as well as difficulties during the fabrication process. To ensure insulation between the $Bi_2Se_3$, b-BP and t-BP and avoid the degradation, few-layer hBN layers with a measured thickness of ~10 nm were capped on to these materials.

We have used a CVD system to synthesize $Bi_2Se_3$ nanoplates on mica substrates. Se powders (0.20 g) (99.9%, Alfa) in a ceramic boat placed at the upstream of the quartz tube furnace (~220 °C). A mica substrate was placed cover the ceramic boat with $Bi_2Se_3$ powder (10 mg) in the center of the furnace. The quartz tube was purged with ultrahigh-purity argon (Ar) gas (99.999%) for 2 min, and then ramped up to 560 °C in 20 min and held at 560 °C for 10 min under a constant flow of Ar (30 sccm) and $H_2$ (10 sccm). After the reaction was completed, the furnace was naturally cooled to ambient temperature, and the aimed materials grown on the mica substrate were collected for subsequent characterizations and manufactures. The $Bi_2Se_3$ nanoplates with measured thickness of ~60 nm, were then transferred to PDMS for the next transfer step on the optical fiber endface.

**EBL process on the fiber endface.** To precisely fabricate the electrode structures on the optical fiber endface, we customized an optical fiber holder to allow the multi-step spin coating processes on the fiber endface. In this process, totally three layers of resist films were coated, consisting of 9% methyl methacrylate (MMA) in ethyl lactate, 6% polymethyl methacrylate (PMMA) in anisole and water-soluble polyaniline (AR-PC 5090) (Supplementary Fig. S4). Compared with PMMA, MMA was more sensitive to the electron beam. As a result, a trapezoidal exposure area was formed on the cross-section of the resist after exposure. The electrodes on the optical fiber endface were then prepared by electron beam lithography (EBL, TESCAN, MIRA3), electron beam evaporation (VZZS-550, VNANO) and lift-off processes.

**Fabrication process of the device.** As shown in Supplementary Fig. S5, after the preparation of electrodes on the fiber endface by EBL, Au/Cr deposition and lift-off processes, $Bi_2Se_3$ and hBN on PDMS were transferred onto the electrodes by dry transfer method. Then the b-BP and t-BP units were fabricated with the same process.

**Measurement configurations.** The experimental setup is demonstrated in Supplementary Fig. S6. An ASE light source was used to provide an infrared laser with 1530–1560 nm wavelength. The laser was coupled to an optical fiber, and a 90:10 optical fiber coupler was used to connect a power meter (S145C, Thorlabs) to monitor the light power. The laser wavelength was fixed at 1550 nm using a filter (FB1550-12, Thorlabs), and the polarization angle of incident light was fixed by a polarizer (FBR-LPNIR, Thorlabs) and changed by rotating a half-wave plate (FBR-AH3, Thorlabs) and a quarter-wave plate (RABQ-1600, Thorlabs), which were both placed in a fiber bench (FB-76W, Thorlabs) with fiber ports (PAF-X-2-C, Thorlabs). The electrode pairs on the optical fiber endface were connected to digital source measure units (Keithley SMU 2450, Tektronix), whose negative electrodes were both grounded.

In the frequency response measurement, a signal generator (33522A, Agilent) and an electro-optic modulator (LN81S-FC, Thorlabs) were used to modulate the light, an amplifier (DLPCA-200, Femto) was used to magnify the photovoltage generated by the BP in self-driven mode, and an oscilloscope (MSOX6004A, Keysight) was used to monitor the high-frequency signal.

Photocurrent mapping measurements were performed by a microscope system (alpha-300R, WITec). The polarimeter sample was fixed in the integrated microscope system, and the incident light was focused on the surface of the sample through the objective lens of the microscope and scanned in the x-y direction. The magnitudes of the device photocurrents corresponding to each x-y coordinate were recorded.

**Single-pixel polarimetric imaging.** We used a 1550 nm laser (VFLS-1550, Connet) and DMD (DLP Discovery 3000, Vialux) to generate Hadamard basis patterns. Light was modulated by a polarizer (GCL-0510, Daheng Optics) and a half-wave plate (GCL-0604, Daheng Optics). Then, the light illuminated the target object and was collected by a fiber coupling system, and the photocurrent signals generated by the twisted BPs were collected and recorded by a source meter (Keithley SMU 2450, Tektronix). For single-pixel polarimetric imaging, Hadamard basis patterns were projected onto the target object, and the polarimetric images were reconstructed by differential HSI technique.

## Acknowledgements

We thank Prof. Peng Chen and Prof. Wei Hu for the helps in the work. This research was sponsored by the National Natural Science Foundation of China (61925502, 51772145, 62135007) and the National Key R&D Program of China (2017YFA0303700, 2018YFA0305800). Natural Science

Foundation of Jiangsu Province (Grant No. BK20180003), 333 high level talent training project of JiangSu.

## Authors' contributions

F.X. and Y.F.H. conceived the project and designed the experiments. Y.F.X., Y.S.W., H.T.X., C.H.W. and Y. M. fabricated and measured the devices performance. R.Z.Z., Y.F.X., Y.L. and Y.C. performed the single-pixel imaging experiments. Y.S.W. and Y.F.X. contributed to material and device characterization. Y.F.X., Y.S.W., and R.Z.Z analyzed the data. X.H.C., M.Z.S. and T.T., K.W. provided BP and hBN crystals, respectively. Y.F.H., F.X. and P. Z. advised on the experiments. Y.F.X., Y.S.W. and R.Z.Z. wrote the original draft. F.X., Y.F.H., P.Z., J.H.C. and Y.Q.L. edited the paper. All the authors reviewed and advised on the paper. F.X. and Y.F.H. guided the research and supervised the project.

## Competing interests

The authors declare no competing interests.

## Additional information